\documentclass[aps,showpacs,nofootinbib,superscriptaddress]{revtex4}
\usepackage{graphicx}
\usepackage{bm}
\usepackage{amsmath}
\usepackage{amssymb}
\usepackage{hyperref}

\newcommand{\qslash}{\kern 0.2 em n\kern -0.50em /}
\newcommand{\nslash}{\kern 0.2 em n\kern -0.50em /}
\newcommand{\kslash}{\kern 0.2 em k\kern -0.45em /}
\newcommand{\lslash}{\kern 0.2 em l\kern -0.50em /}
\newcommand{\pslash}{\kern 0.2 em p\kern -0.50em /}
\newcommand{\Sslash}{\kern 0.2 em S\kern -0.50em /}
\newcommand{\Pslash}{\kern 0.2 em P\kern -0.50em /}
\newcommand{\Dslash}{\kern 0.2 em D\kern -0.65em /\kern 0.15em}

\usepackage{color}

\begin{document}
\title{The electromagnetic form factors of $\Lambda$ hyperon in $e^+e^-\rightarrow \bar\Lambda\Lambda$}
\author{Yongliang Yang}\affiliation{School of Physics, Southeast University, Nanjing
211189, China}
\author{Zhun Lu}\email{zhunlu@seu.edu.cn}\affiliation{School of Physics, Southeast University, Nanjing
211189, China}

\begin{abstract}
  We study the electromagnetic form factors of $\Lambda$ hyperon in the timelike region using the recent experimental data in the exclusive production of $\bar{\Lambda} \Lambda$ pair in electron-position annihilation. We present a pQCD inspired parametrization of $G_E(s)$ and $G_M(s)$ with only two parameters, and we consider a suppression mechanism of the electric form factor $G_E(s)$ compared to the magnetic form factor $G_M(s)$. The parameters are determined through fitting our parametrization to the effective form factor data in the reaction $e^+e^-\rightarrow \bar\Lambda\Lambda$. Except the threshold region, our parametrization can reproduce satisfactorily the known behavior of the existing data from the BaBar, DM2 and BESIII Collaborations. We also predict the double spin polarization observables $A_{xx}$, $A_{yy}$ and $A_{zz}$ in $e^+e^-\rightarrow \bar\Lambda\Lambda$.
\end{abstract}


\maketitle
\section{introduction}
\label{sec:intro}

The electromagnetic form factors (EMFFs) of hadrons are important quantities for probing the fundamental information of the hadron structure and understanding the strong interaction in the perturbative and nonperturbative region.
The investigations on the proton and neutron form factors have been performed extensively in both the spacelike and timelike regions, i.e. in the $ep$ elastic scattering, $\bar{p}p$ annihilation and $e^+e^-$ annihilation processes.
Particularly, the Born cross section of nucleon pair production and the corresponding nucleon effective form factor were studied theoretically and experimentally over the last two decades~\cite{Delcourt:1979ed,Bisello:1990rf,Armstrong:1992wq,Bardin:1994am,Antonelli:1998fv,Kuraev:2011vq,
Andreotti:2003bt,Brodsky:2003gs,Aubert:2005cb,Pedlar:2005sj,TomasiGustafsson:2005kc,
Denig:2012by,Lees:2013ebn,Achasov:2014ncd,Haidenbauer:2014kja,Akhmetshin:2015ifg,Ablikim:2015vga}.

In recent years, there is also an increasing interest on the EMFFs of another baryon, i.e. the $\Lambda$ hyperon.
In contrast to nucleons, it is rather difficult to explore the cross section and the EMFFs of the $\Lambda$ hyperon~\cite{Ablikim:2017pyl,Dobbs:2014ifa,Aubert:2007uf,Bisello:1990rf}.
This is because hyperons are unstable and hyperon targets are unfeasible, in principle the EMFFs of hyperons in the spacelike cannot be measured by exclusive experiments.
Therefore, the timelike form factors can offer a unique opportunity to study the electromagnetic property of hyperons.
The importance of studying hyperon structure and measuring the timelike form factors of the $\Lambda$ hyperon was first indicated by Cabibbo and Gatto~\cite{Cabibbo:1960zza}.
The BaBar~\cite{Aubert:2007uf} and DM2~\cite{Bisello:1990rf} Collaboration experiments measured the Born cross section and effective form factor of the process $e^+e^-\rightarrow\bar\Lambda \Lambda$, with significantly lager uncertainties compared to the proton case. Very recently, BESIII~\cite{Ablikim:2017pyl} Collaboration also provide new measurement on this process, and the result is found to be consistent with the previous measurement with improved precision, particularly in the near threshold region.
Theoretically, the near threshold behaviour of the baryon pair production has been investigated by several theoretical works~\cite{Haidenbauer:2016won,ElBennich:2008vk,Haidenbauer:2006dm,Fonvieille:2009px,Dalkarov:2009yf}.
The EMFFs of the $\Lambda$ hyperon in the timelike region have been analyzed by several theoretical studies~\cite{Haidenbauer:2016won,Faldt:2017kgy,Faldt:2016qee,Faldt:2013gka,
Dalkarov:2009yf,Baldini:2007qg}.
Particularly, the first attempt to calculate the $\Lambda$ hyperon EMFFs $G_E$ and $G_M$ up to the $\bar{\Sigma}^+\Sigma^+$ threshold was reported in Ref.~\cite{Haidenbauer:2016won}.
Furthermore, at large center-of-mass (c.m.) energy $\sqrt{s}$, perturbative QCD predicts the existence of final interactions in the reaction $e^+ e^- \rightarrow \bar{B} B $, since the hadrons are initially produced with small color dipole moment~\cite{Brodsky:2003gs}.
These studies on the existing data and the future plans of precise measurements of the exclusive production of $\Lambda$ hyperons provide useful constraints on the $\Lambda$ electromagnetic form factor in the timelike region.

In this work, we suggest a parametrization on the lambda EMFFs with only two parameters to describe the existing data of the lambda EMFFs as well as the Born cross section of the reaction $e^+e^-\to \bar{\Lambda}\Lambda$, up to $\sqrt{s} =3.08$ GeV.
The parameterized form is based on the quark counting rules within pQCD and a suppression mechanism of the electric form factor~\cite{Kuraev:2011vq}.
Similarly to the nucleon, we take an assumption that the magnetic form factor $G_M$ of the $\Lambda$ hyperon is dominant for the contribution in $e^+e^-\rightarrow \bar\Lambda \Lambda$~\cite{TomasiGustafsson:2005kc}.
The parametrization provides a good description for the cross section of $e^+e^-\to \bar{\Lambda}\Lambda$ and the corresponding EMFFs data from the DM2, BaBar and BESIII experiments in a wide range of $\sqrt{s}$, except the region near the $\bar\Lambda \Lambda$ threshold.
In Addition, we predict spin polarized observables in $\Lambda$ pair production which are related to the moduli of $G_E$ and $G_M$.


\section{Basic formalism for $e^+e^-\rightarrow \bar\Lambda\Lambda$}
\label{formalism}

The general expression of the Born cross section for the reaction $e^+e^-\rightarrow \bar B B$ has been given in Ref.~\cite{Haidenbauer:2014kja}, where $B$ is a spin-$1/2$ baryon.
Under the one-photon exchange approximation, the angular dependent differential cross section of the lambda pair production is governed by the electric and magnetic form factors $G_E$ and $G_M$ as follows
\begin{align}
{d\sigma(s)\over d\Omega}={\alpha^2\beta C\over 4s}\bigg{[}|G_M(s)|^2(1+\cos^2\theta)+{1\over 2\tau}|G_E(s)|\sin^2\theta\bigg{]}\,,
 \label{eq:ddis}
\end{align}
with $\tau={s/4M^2}$, where $M=1.116$ GeV is the lambda mass.
Here, $\alpha\approx 1/137$ is the electromagnetic fine-structure constant, $\theta$ is the scattering angle in the c.m. frame, $s=q^2$ is the square of the c.m. energy.
The variable $\beta={k_\Lambda\over k_e}$ is a phase-space factor, with $k_\Lambda $ and $k_e$ the of the three-vector momenta of the lambda and the electron in the c.m. frame~\cite{Haidenbauer:2016won}, respectively.
In the differential cross section, we can obtain the phase factor $\beta=\sqrt{1-1/\tau}$ by setting the lepton mass to zero.
The Coulomb enhancement factor $C$~\cite{Baldini:2007qg,Arbuzov:2011ff}, accounting for the electromagnetic interaction of the point-like baryon pairs in the final state, is equal to one for neutral baryon pair and $y/(1-e^{-y})$ for charged baryon pair~\cite{Tzara:1970ne} with $y=\pi\alpha(1+\beta^2)/\beta$ .

The total cross section thus can be expressed in terms of $G_{M/E}$
\begin{align}
\sigma(s)={4\pi\alpha^2\beta\over 3s}\bigg{[}|G_M(s)|^2+{1\over 2\tau}|G_E(s)|\bigg{]}\,.
 \label{eq:sidis}
\end{align}
Another quantity used in various analyses is the so-called the effective form factor $G_{\text{eff}}(s)$.
The effective form factor is equivalent to $|G_M|$ under the working hypothesis $G_E=G_M$~\cite{TomasiGustafsson:2005kc}.
In more general cases, the effective form factor is relevant to the combination of the moduli of the EMFFs~\cite{Aubert:2005cb}
\begin{align}\label{Geff}
|G_{\text{eff}}(s)|=\sqrt{2\tau|G_M(s)|^2+|G_E(s)|^2\over 1+2\tau}\,,
\end{align}
it is proportional to the square root of the Born cross section via
\begin{align}\label{Geff}
|G_{\text{eff}}(s)|&=\sqrt{\sigma_{e^+e^-\rightarrow \bar\Lambda\Lambda}(s)\over {4\pi\alpha^2\beta\over 3s}C[1+{1\over 2\tau}]}\,,
\end{align}
which can be deduced from the experimental measurements.
In Ref.~\cite{Haidenbauer:2016won}, a variety
of $\Lambda \bar{\Lambda}$ potential models~\cite{Haidenbauer:1991kt,Haidenbauer:1992wp} is employed
to estimate the effective form factor, as well as the electromagnetic form factors $G_M(s)$ and $G_E(s)$, with $s$ up to the $\bar{\Sigma}^+ \Sigma^+$ threshold.

\section{The parametrization and fitting procedure in the timelike region}
\label{data}

The investigation of the EMFFs in the time like region can be accessed from the EMFFs in the spacelike region via the crossing symmetry between the electron-nucleon elastic scattering and the
annihilation processes~\cite{Denig:2012by}.
For electron-nucleon elastic scattering, the reaction process of $\gamma^*+3q\rightarrow 3q$ in pQCD is described by the Feynman diagram illustrated in Ref.~\cite{Pacetti:2015iqa},
where the virtual photon interacts with the proton leaving the proton unchanged after transferring the momentum to each of the quarks, through the gluon exchange~\cite{Lepage:1980fj,Lepage:1979za}.
Based on the quark counting rules, pQCD predicts~\cite{Pacetti:2015iqa} that the nucleon EMFFs behave as $G_{E(M)}~\sim {(-q^2)}^{1-n_h}$ at large momentum transfer, where $n_h$ is the number of the valance quarks in the baryons~\cite{Matveev:1973uz,Pacetti:2015iqa}, and $n_h-1$ is the number of the exchanged gluons between the quarks.
Hence the nucleon EMFFs in the spacelike region must contain terms with, at least, two gluon propagators that entail the power law behavior as
\begin{align}\label{form}
  |G_M|&\sim \bigg{(}{1\over -q^2}\bigg{)}^2\,,
\end{align}
where $q$ is the transferred momentum in the process.

Based on the pQCD predictions and counting rules~\cite{TomasiGustafsson:2005kc}, we present a parametrization for the lambda $|G_M|$ and $|G_E|$ in the timelike region as follows
\begin{align}
  |G_M(s)|&={A_\Lambda\over{\tau^{2+\delta_\Lambda}\ln^2(s/\Lambda^2_{QCD})}}\,, \label{eq:gegm}\\
  |G_E(s)|&=\tau^{-1}|G_M(s)|\,, \label{eq:ge}
\end{align}
where $\Lambda_{QCD}=0.3~\text{GeV}$~\cite{TomasiGustafsson:2005kc} is the QCD scale parameter, $A_\Lambda$ and $\delta_\Lambda$ are the two free parameters which can be obtained by fitting the experimental data.
The variable $\tau$ is introduced in Eq.~(\ref{eq:gegm}) to make $A_\Lambda$ dimensionless.
The factor $\ln(s/\Lambda^2_{QCD})$ represents the logarithmic corrections from QCD.
As shown in Refs.~\cite{Brodsky:2003gs,Sudol:2009vc,TomasiGustafsson:2005kc,Brodsky:2003pw,Belitsky:2002kj} logarithmic corrections enable a good fit to the data.

The main difference between our parametrization for Lambda EMFFs and the one for the nucleon EMFFs in spacelike region~\cite{TomasiGustafsson:2005kc} is their power law behavior.
In the case of the nucleon EMFFs, the power of $1/(-q^2)$ is 2, representing the minimal number of the exchanged gluons~\cite{TomasiGustafsson:2005kc,Pacetti:2015iqa}.
In our case we introduce the parameter $\delta_\Lambda$ accounting the difference between the $\Lambda$ hyperon and the nucleons as well as the difference between the timelike and spacelike region, that is, here $2+\delta_\Lambda$ gives the averaged number of the exchanged gluons for the $\bar{\Lambda} \Lambda$ production, beyond the minimal gluon exchange.
Secondly, ${4M^2/s}$ in Eq.~(\ref{eq:ge}) severs as an additional suppression factor for $G_E$ compared to $G_M$, due to the screening of electric charge in a neutral plasma~\cite{Kuraev:2011vq}.
The similar suppression effect was already considered in the study of the nucleon EMFFS~\cite{Kuraev:2011vq}.   Finally, our parameterization is consistent to the normalization condition $|G_E(4M_\Lambda^2)/G_M(4M_\Lambda^2)|=1$ at the kinematical threshold.

\begin{figure}
  \includegraphics[width=0.48\columnwidth]{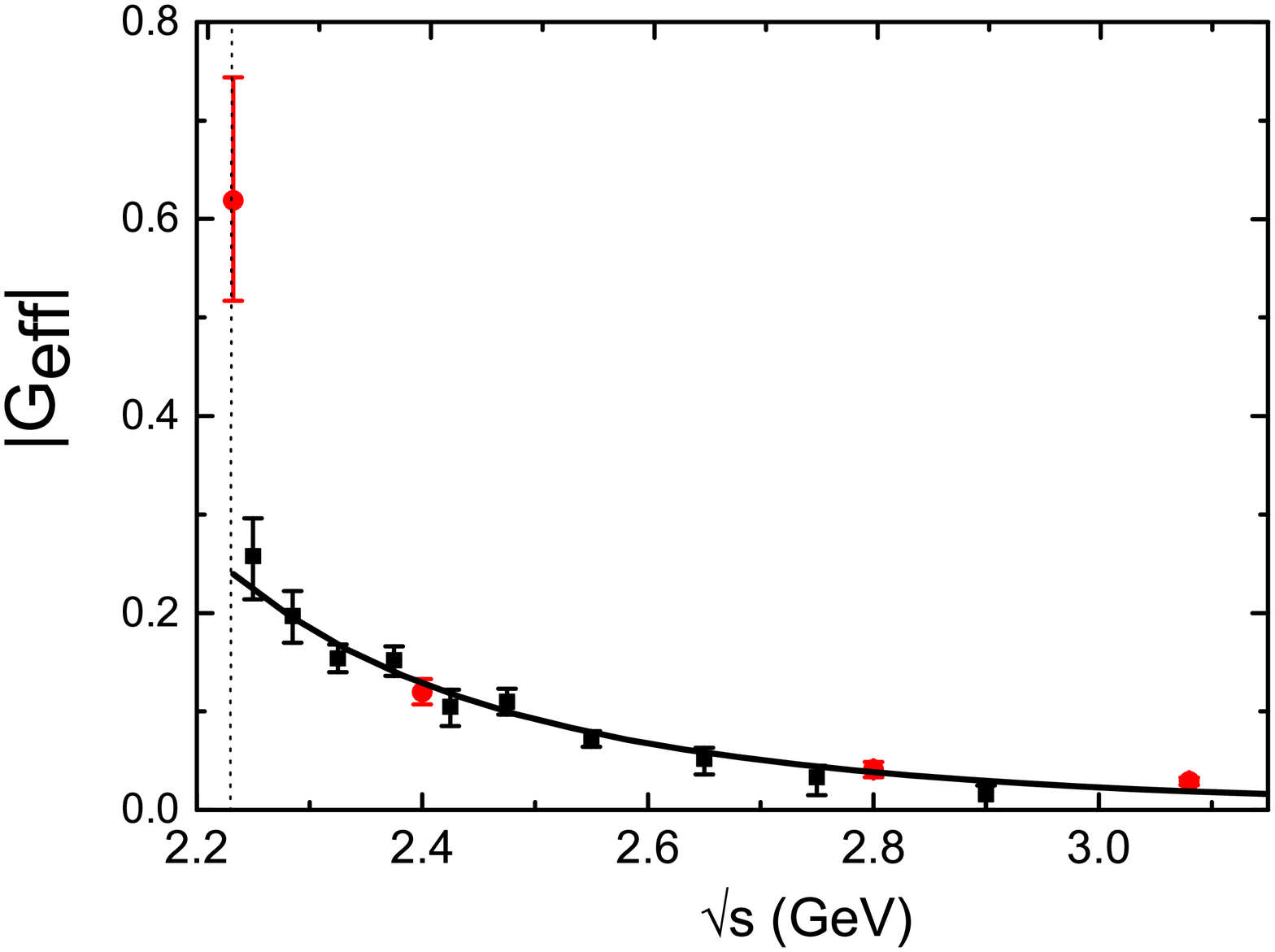}
  \includegraphics[width=0.48\columnwidth]{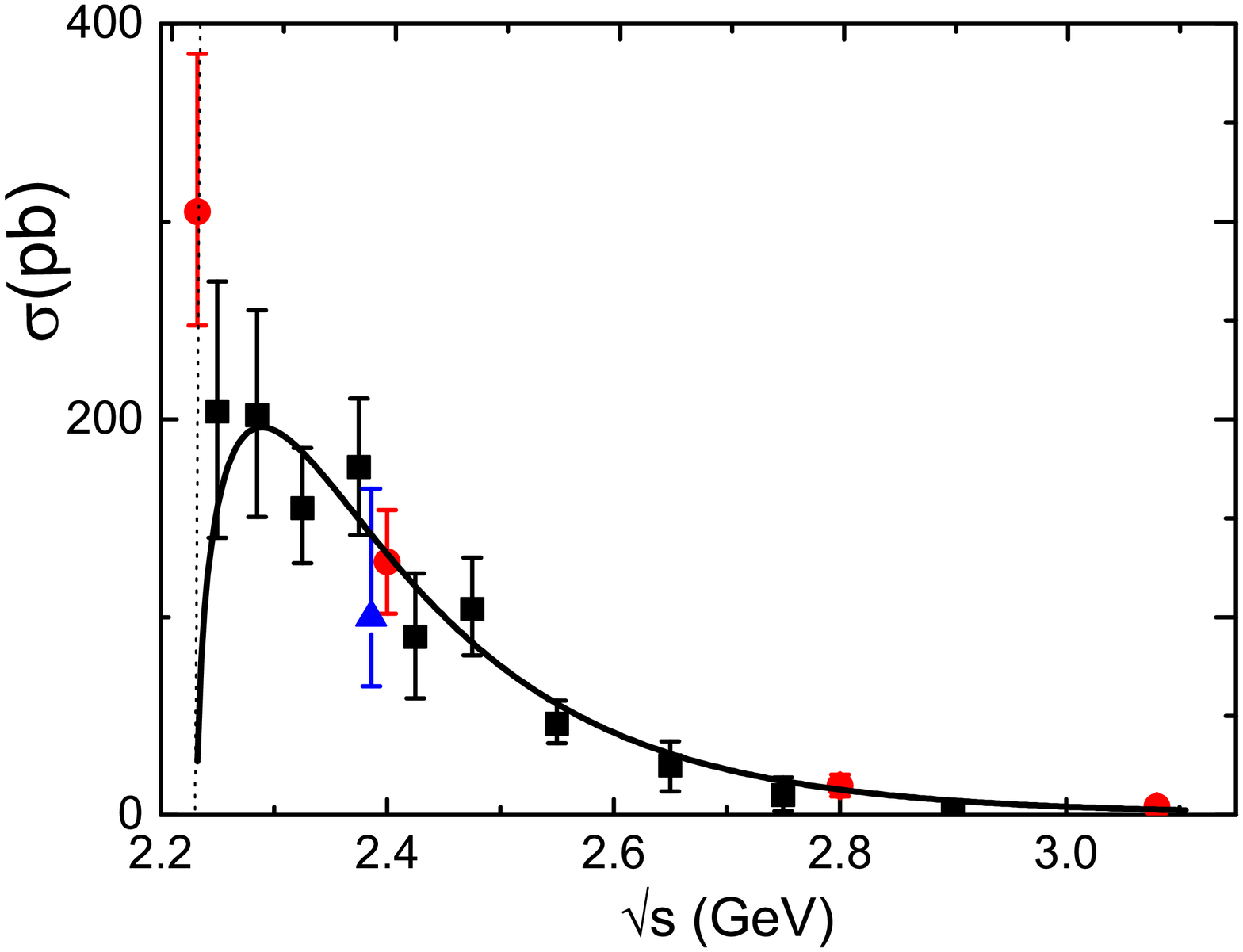}
  \caption{Left panel: our fit to the world data on the lambda $|G_{\text{eff}}|$. Right panel: the model resulting Born cross section $\sigma$ vs. $\sqrt{s}$ compared with data.
  The rectangles, triangles and circles  represent the data from the BaBar~\cite{Aubert:2007uf}, DM2~\cite{Bisello:1990rf} and BESIII~\cite{Ablikim:2017pyl} Collaborations, respectively. The vertical lines represent the $\bar{\Lambda}\Lambda$ threshold.}
 \label{gsigma}
\end{figure}

The Born cross section for the reaction $e^+ \,e^-\to \bar{\Lambda}\,\Lambda $ has been measured by BaBar~\cite{Aubert:2007uf}, DM2~\cite{Bisello:1990rf} and BESIII~\cite{Ablikim:2017tys,Ablikim:2017pyl}, covering the mass region from $\sqrt{s}=2.2324~\text{GeV}$ to $\sqrt{s}=3.08~\text{GeV}$ .
We fit the data of the lambda effective form factor $|G_{\text{eff}}|$ by using the parameterization in Eqs.~(\ref{eq:gegm}) and (\ref{eq:ge}).
The best fitted values of the parameters are
\begin{align}
A_\Lambda=3.781, ~~~ \delta_\Lambda=1.362.
\end{align}
In the left panel of Fig.~\ref{gsigma}, the comparison between our fitting and the experimental measurement on $G_{\text{eff}}$ is depicted. In the right panel of Fig.~1, we plot the estimated Born cross section in reaction $e^+ e^- \to \bar{\Lambda} \Lambda$ together with the BaBar~\cite{Aubert:2007uf}, DM2~\cite{Bisello:1990rf} and BESIII data.
We find that except the near-threshold region, our parametrization can well describe the data of $G_\text{eff}$ as well as the cross section.
The underestimate of our model in the threshold region is understandable since our model is based on the perturbative QCD.
The unusual threshold behaviour implies a more complicated underlying physics scenario other than perturbative QCD.
Interpretations on the near threshold enhancement of the cross section were suggested in Ref.~\cite{Baldini:2007qg,Haidenbauer:2016won}.

\begin{figure}
  \includegraphics[width=0.4\columnwidth]{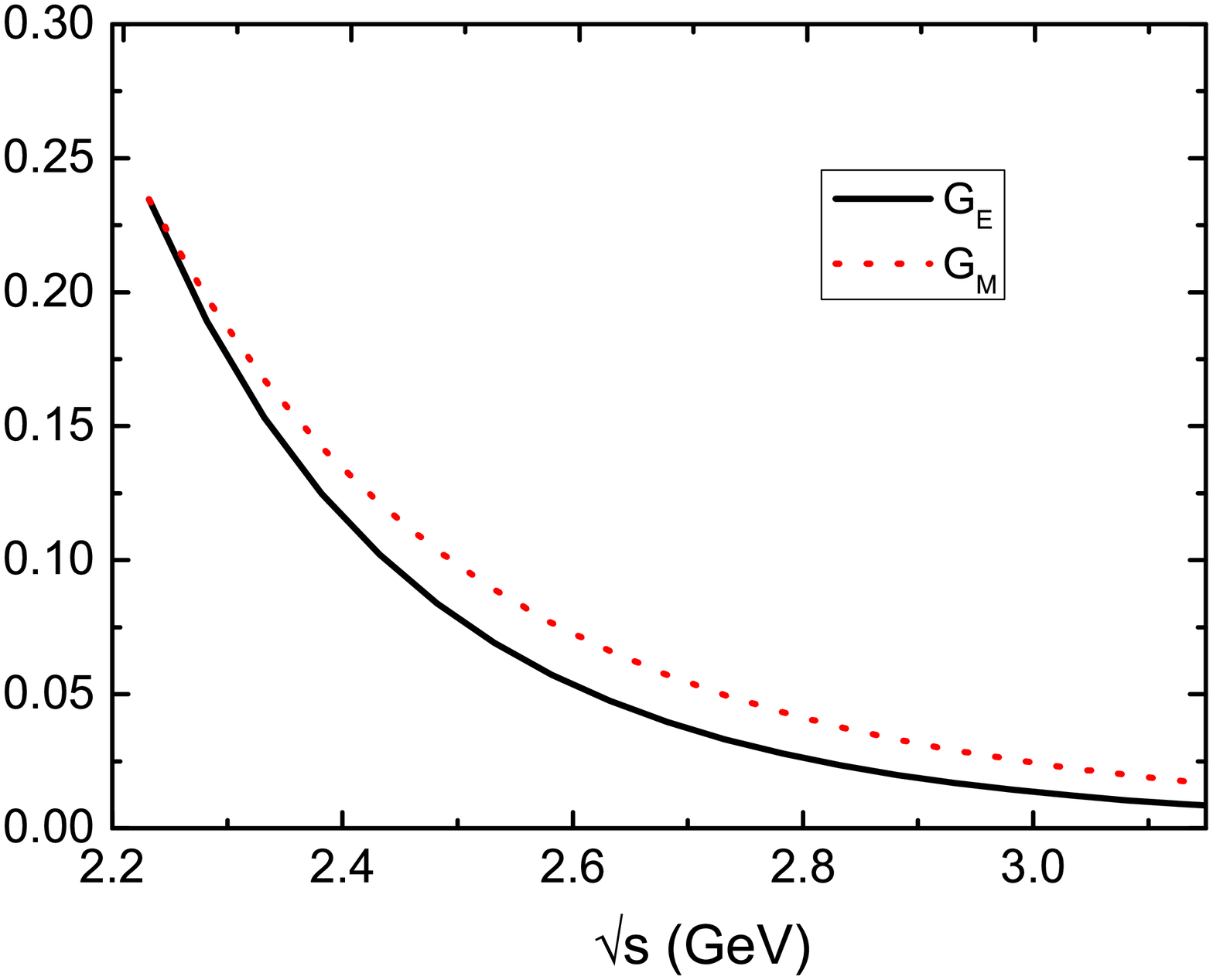}
  \includegraphics[width=0.4\columnwidth]{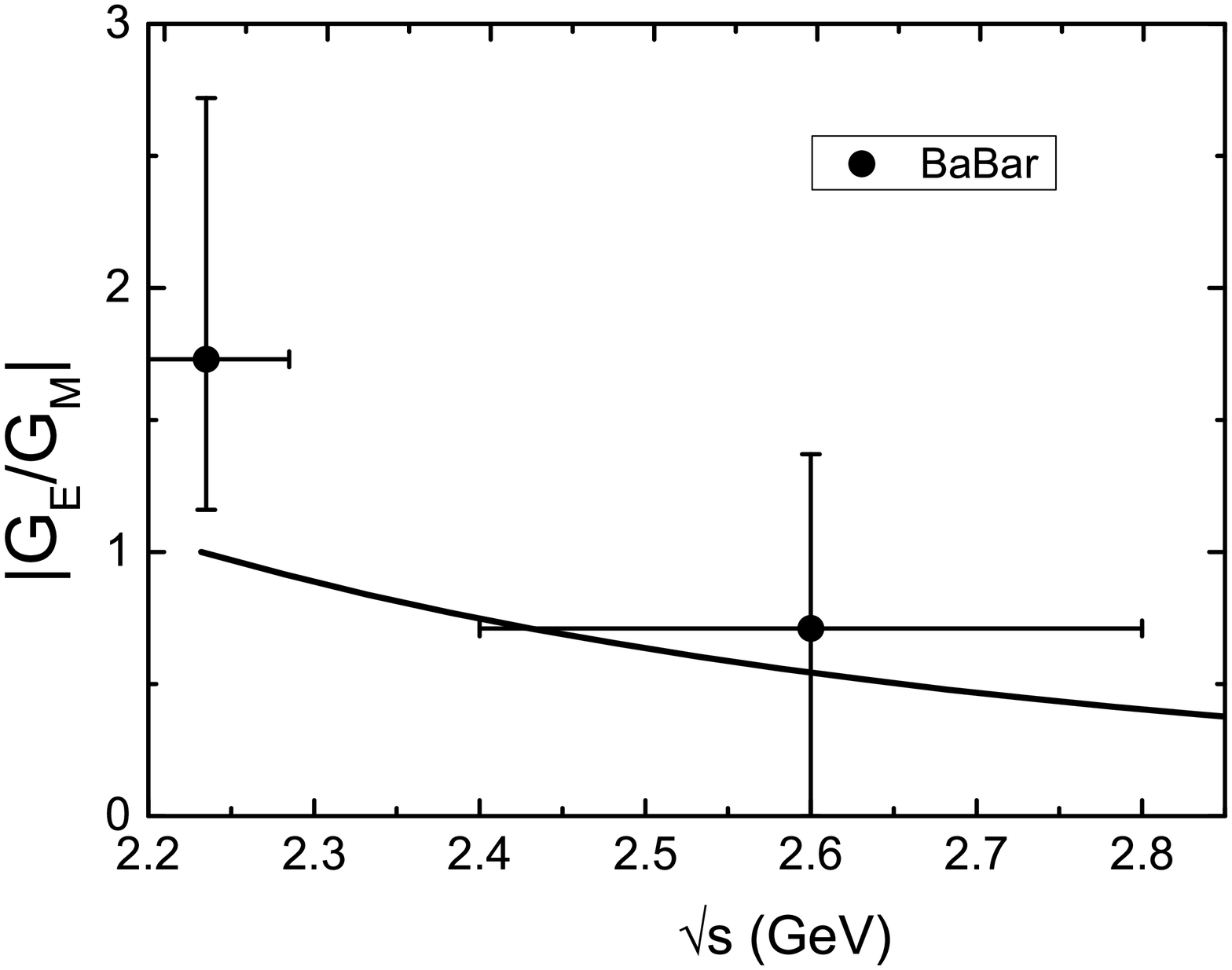}
  \caption {Left panel: the the electric form factor $|G_E|$ (solid line) and the magnetic form factor $|G_M|$ (dashed line) of the lambda in our model, as functions of $\sqrt{s}$. Right panel:
  the ratio $|G_E/G_M|$ of the lambda. Data are from the BaBar Collaboration~\cite{Aubert:2007uf}.}
 \label{ratio}
\end{figure}

In the left panel of Fig.~\ref{ratio}, we plot the moduli of the lambda $G_E$ (solid line) and $G_M$ (dashed line), respectively. Due to our parametrization in Eq.~(\ref{eq:ge}), the electric form factor is smaller than $G_M$ in the region $s> 4M^2$.
Although at present there is no individual experimental measurements on $G_E$ and $G_M$, the ratio of these two quantities, $|G_E/G_M|$, has been measured by the BaBar Collaboration ~\cite{Aubert:2007uf} in two mass intervals: from $\bar\Lambda\Lambda$ threshold to $2.4~\text{GeV}$ and from $2.4~\text{GeV}$ to $2.8~\text{GeV}$.
We plot the model result of the ratio $|G_E/G_M|$ in the right panel of Fig.~\ref{ratio}.
We find our result is in agreement with the data of BaBar at the high mass interval.
The ratio $|{G_E/ G_M}|$ is found to be greater than 1 near-threshold in the measurement, which may be explained by the final state interaction between the baryon pairs ~\cite{Iachello:2004aq,Iachello:2003ep,Haidenbauer:2016won}.

\begin{figure}

  \includegraphics[width=0.32\columnwidth]{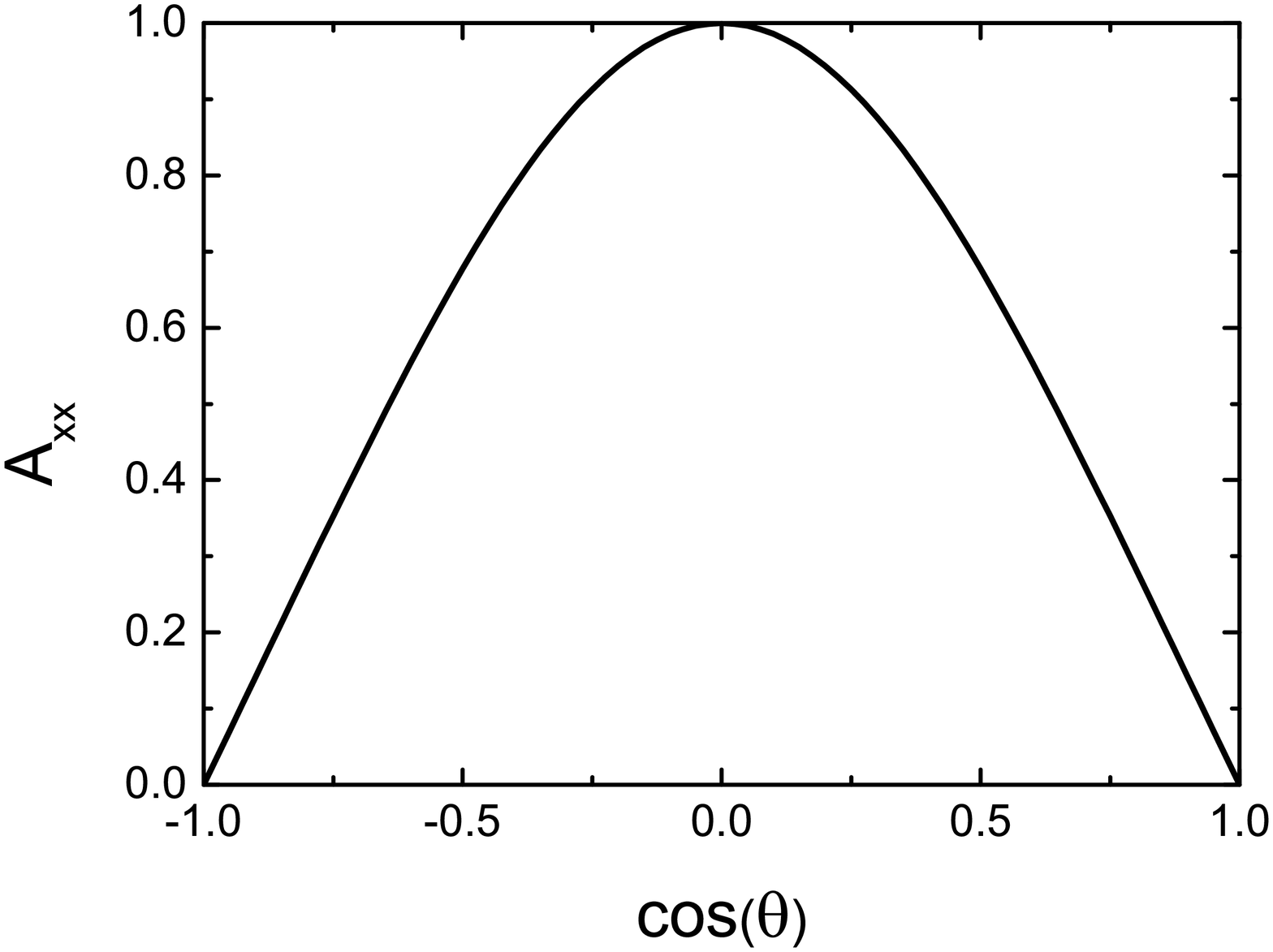}
  \includegraphics[width=0.32\columnwidth]{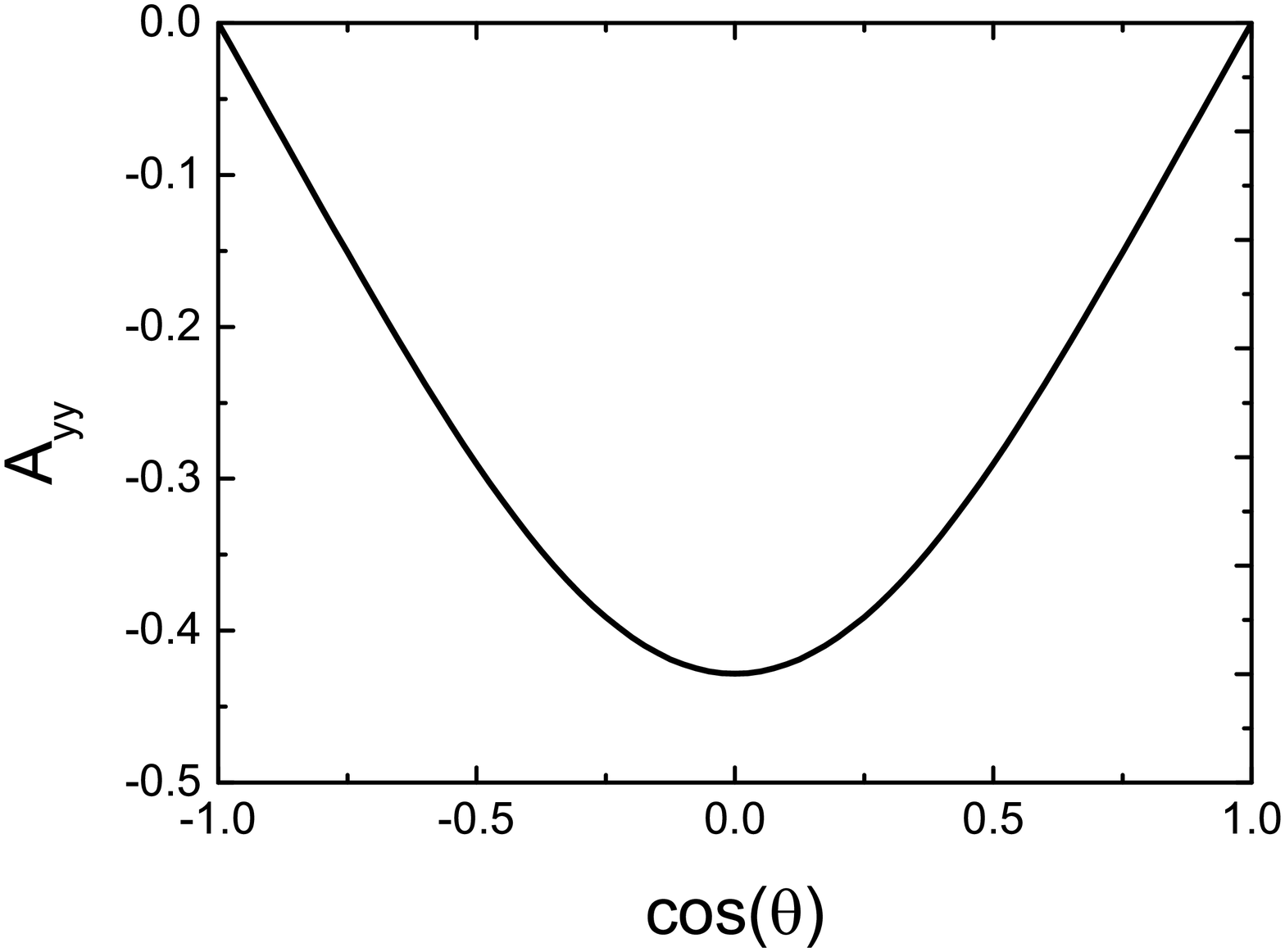}
  \includegraphics[width=0.32\columnwidth]{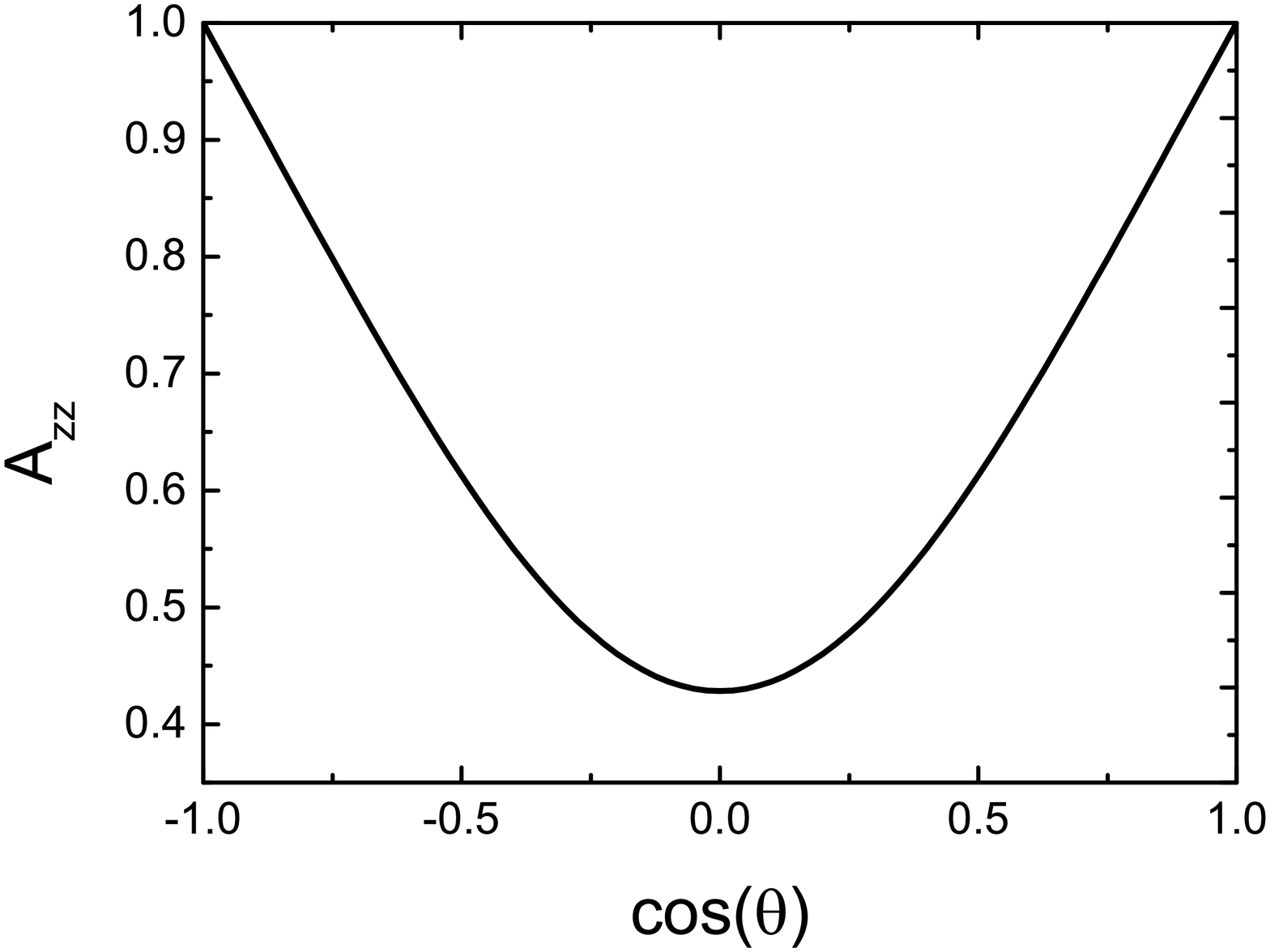}
  \caption {The predictions of the double spin polarization observables vs.  $\cos(\theta)$ in the reaction $e^+e^-\rightarrow \bar\Lambda\Lambda$ at the energy $\sqrt{s}$ = 2.6~\text{GeV}. We calculate these observables in a coordinate system following Ref.~\cite{Faldt:2016qee}. }
  \label{double}
 \end{figure}

The lambda EMFFs can be used to predict the spin polarized observables $A_{ij}$~\cite{Faldt:2017kgy,Faldt:2016qee} appearing in the reaction $e^+ e^- \to \bar{\Lambda}\Lambda$, which depend not only on the moduli of the EMFFs, but also on the relative phase between $G_E$ and $G_M$.
Those observables have not been measured in experiments yet.
As our parametrization cannot separate the real and imaginary parts of the EMFFs, we only calculate the three diagonal components of the spin polarization observables~\cite{Faldt:2016qee}:
\begin{align}
  A_{xx}&={[D_c-D_s]\sin^2(\theta)\over{D_c-D_s\sin^2(\theta)}}\,, \\
  A_{yy}&={-D_s\sin^2(\theta)\over{D_c-D_s\sin^2(\theta)}}\,, \\
  A_{zz}&={[D_s\sin^2(\theta)+D_c\cos^2(\theta)]\over{D_c-D_s\sin^2(\theta)}}\,,
\end{align}
where $\theta$ is the scattering angle defined in the previous section, $D_c=2s|G_M|^2$ and $D_s=s|G_M|^2-4M^2|G_E|^2$.
In our estimate these polarization observables are calculated in the orthonormalized coordinates system~\cite{Faldt:2016qee}.
In Fig.~\ref{double}, we plot our prediction on $A_{xx}$, $A_{yy}$ and $A_{zz}$ at the energy $\sqrt{s}=2.6~\text{GeV}$, where the ratio $|G_E/G_M|$ from our model agrees with the BaBar data.
We find that in our model $A_{xx}$ and $A_{zz}$ are positive in the region $\sqrt{s}=2.6~\text{GeV}$, while $A_{yy}$ is negative in the same region.
The angular dependence of $A_{xx}$ is different from those of $A_{yy}$ and $A_{zz}$.
As the polarization observables are sensitive to the model assumption of the form factors, precise measurements of these quantities would provide a way to test the validity of the EMFFs in different models.

\section{conclusions}
\label{summary}

In this paper, we have generalized the parameterization of the nucleon EMFFs $|G_E|$ and $|G_M|$ to the case of the $\Lambda$ hyperon with only two parameters.
A power law behavior $s^{2+\delta_\Lambda}$ for the EMFFs in the timelike region has been introduced following the quark counting rule inspired by pQCD, with $\delta_\Lambda$ representing the difference between the lambda form factors and the nucleon form factors.
The fit was performed on the existing form factor data from the BaBar, DM2 and BESIII experiments.
The fitted values of the parameters in Eq.~(\ref{eq:gegm}) are $A_\Lambda=3.781$ and $\delta_\Lambda=1.362$, respectively.
The resulting parameters are applied to calculate the Born cross sections in the reaction $e^+e^-\rightarrow \bar\Lambda\Lambda$ as well as the ratio $|G_E/G_M|$.
Except the threshold region $\sqrt{s}=2.2324~\text{GeV}$, our parametrization results are in good agreement with experimental data in a wide $\sqrt{s}$ region.
At last, we have presented our prediction on the diagonal spin polarization observables $A_{xx}$, $A_{yy}$ and $A_{zz}$ at $\sqrt{s} =2.6~\text{GeV}$.
We expect future precise measurement on the cross section and the spin polarized observables in $e^+e^-\rightarrow \bar\Lambda\Lambda$ will shed more light on the size of the lambda EMFFs as well as their phase angles.

\section{Acknowledgements}

This work is partially supported by the NSFC (China) grant 11575043, and by the Fundamental Research Funds for the Central Universities of China.
Y.~Y. is supported by the Scientific Research Foundation of Graduate
School of Southeast University (Grant No. YBJJ1770) and the Postgraduate Research \& Practice Innovation Program of Jiangsu Province (Grants No. KYCX17\_0043).

\end{document}